\documentclass[aps,prb,twocolumn,groupedaddress,showpacs]{revtex4}

\usepackage{graphicx}

\begin{document}


  \title{Electron transport through antidot superlattices in $Si/SiGe$ heterostructures:
new magnetoresistance resonances
 in lattices with large diameter antidots.}



  \author{E.~B.~Olshanetsky$^{1}$, V.~T.~Renard$^{2,3,\footnote{Present address : NTT Basic Research Laboratories, 3-1 Morinosato
Wakamiya, Atsugi 243-0198, Japan}}$, Z.~D.~Kvon$^{1}$,
J~.C.~Portal$^{2,3,4}$,J.~M.~Hartmann$^{5}$}
\affiliation{$^1$ Institute of Semiconductor Physics, Novosibirsk
630090, Russia;}\affiliation{$^2$ GHMFL, MPI-FKF/CNRS, BP-166,
F-38042, Grenoble Cedex9, France;} \affiliation{$^3$ INSA-Toulouse,
31077, Cedex 4, France;}\affiliation{$^4$ Institut Universitaire de
France, Toulouse, France;}\affiliation{$^5$ CEA/Leti, F-38054,
Grenoble, Cedex 9, France}

\date{20 September 2006}

\begin{abstract}
  In the present work we have investigated the transport properties in a
number of Si/SiGe samples with square antidot lattices of different periods. In
samples with lattice periods equal to 700 nm and 850 nm we have observed the
conventional low-field commensurability magnetoresistance peaks
consistent with the previous observations in GaAs/AlGaAs and Si/SiGe samples
with antidot lattices. In samples with a 600 nm lattice period a new
series of well-developed magnetoresistance oscillations has been found
beyond the last commensurability peak which are supposed to originate from
periodic skipping orbits encircling an antidot with a particular number of
bounds.

\end{abstract}

\maketitle

\section{Introduction}
  A two dimensional (2D) system modulated by a periodic strong repulsive
potential is called an antidot lattice. Over the last 15 years various
interesting phenomena have been observed in antidot lattices in uniform
perpendicular magnetic fields. They are the low-field resonances in the
longitudinal magnetoresistance, the corresponding non-quantized Hall
plateaus and the quenching of the Hall effect near zero magnetic field,
\cite{Enslin,Weiss,Lorke,Schuster,G.M.Gusev,G.M.Gusev1}. These effects can be
interpreted in terms of classical cyclotron orbits that are commensurate
with the antidot lattice. A more thorough classical analysis based on the
Kubo formula reveals that these effects have their origin in the magnetic
field dependent mixture of chaotic and regular trajectories,
\cite{Baskin,Fleischmann}.

  Originally, all the antidot lattice related effects were observed in the
interval between zero magnetic field and the field at which the last
commensurability resonance $2r_c=a$ occurs, where $r_c$ is the cyclotron
radius and $a$ is the antidot lattice period. Recently, however , it has
become clear that under certain conditions a new type of longitudinal
magnetoreistance resonances can be observed in antidot lattices at magnetic
fields following the main commensurabilty peak $2r_c=a$ and before the
beginning of the Shubnikov-de Haas oscillations, \cite{M.V.Budantsev,Eroms,Z.D.Kvon}.
Although the interpretation of these resonances
is not yet fully developed, it is already clear that a large antidot
diameter is necessary for their observation.

  Experiments in antidot lattices usually require that the elastic mean free
path of the electrons should be larger than the superlattice period. This
explains why previously the majority of such experiments were carried out in
high mobilty GaAs/AlGaAs heterostructures. Now, the progressively improving
quality of the 2D electron gas in Si/SiGe heterostructures has made it an
equally suitable material for the studies of electron transport in antidot
lattices, ~\cite{Tobben}.

  In the present work we have investigated the transport properties of a
number of square antidot lattices with different periods fabricated on top
of a high mobility Si/SiGe heterostructure. In the samples with lattice
periods 850 and 700 nm we observe the usual commensurability peaks and the
shoulder-like feature at a higher field, already reported in ~\cite{Tobben}.
In the samples with the smallest lattice period (600 nm) a number of well
developed and high-amplitude magnetoresistance oscillations are found in
magnetic fields following the main commensurability peak $2r_c=a$. These new
oscillations differ in certain respects from those reported in
\cite{M.V.Budantsev,Eroms,Z.D.Kvon} and are superior to them in size.

\section{Experimental}

  Our samples were Hall bars fabricated on top of a
$Si/Si_{0.75}Ge_{0.25}$ heterostructure with a high mobility 2D electron
gas, ~\cite{Hartmann}. The distance between the potential probes was $100$~$\mu$m; the Hall bar
width was $50$~$\mu$m. The parameters of the original heterostructure were as follows:
the electron density $N_s=(5.8-6)\times10^{11}$cm$^{-2}$, the mobility
$\mu=(1.6-2)\times 10^{5}$cm$^2$/Vs. A square array of antidots with a lithographical
diameter $d_l=150-200$~nm, fabricated by electron beam lithography and
reactive plasma etching covered the whole segment of the samples between the
voltage probes. The total number of antidots was $(7-12)\times10^3$. We
have investigated six samples, two of each type, with different lattice
periods equal to $600$, $700$ and $850$~nm. The magnetoresistance was measured using
a conventional four point ac lock-in scheme in a $He_3$ cryostat at
temperatures $0.3-4.2$~K and in magnetic fields up to $13$ ~T.

\section{Results and discussion}

  Fig.1 shows a typical magnetoresistance dependence for sample $D18-01-10$
with a 700 nm antidot lattice period. At low fields two commensurability
peaks in the magnetoresistance curve can be observed. These well-known
maxima were also found in the samples with the lattice period equal to 850 nm.
Following these maxima there is a shoulder-like feature. At still higher
fields Shubnikov-de Haas oscillations develop that can be used to determine
the sheet electron density: $N_s=5.3\times10^{11}$cm$^{-2}$.

The positions of the low-field peaks are $(\bf 1)$ - $B=0.058$~T
($r_c/a\approx1.75$) and $(\bf 2)$ - $B=0.233$~T
($r_c/a\approx0.53$) respectively. Such peaks are also commonly
observed in similar in size square antidot lattices on
$AlGaAs/GaAs$, ~\cite{Lorke,Schuster,G.M.Gusev,G.M.Gusev1}. Both the
position and the magnitude of these peaks can be satisfactorily
described by a model that takes into account two special types of
regular electron trajectories in an antidot lattice. Namely, this
model considers the contribution, on the one hand, of the so-called
"pin-ball" trajectories localized around 1, 4 and more antidots
(Ref.~\cite{Weiss}) and, on the other hand, of the delocalized
"runaway" trajectories that bounce away along a row of neighboring
antidots, Ref.~\cite{G.M.Gusev1}. It is established that the
relative contribution of the "pin-ball" and the "runaway"
trajectories to the formation of the main commensurability peaks
depends on the ratio $d_e/a$, where $d_e$ is the antidot effective
diameter and $a$ is the lattice period. So, for $d_e/a<0.5$ the
contribution of the "runaway" trajectories dominates, while for
$d_e/a \leq 1$ it is the "pin-ball" trajectories that give rise to
the main commensurability peaks. In our case, peak $\bf 1$ in Fig.1
must be due to "runaway" trajectries since in a lattice with the
parameters of sample $D18-01-10$ a "pin-ball" trajectroy cannot give
rise to a peak at the corresponding magnetic field.

The shoulder at $B=0.376$~T ($r_c/a\approx0.33$), that has also been observed in
Ref.~\cite{Tobben}, will be discussed below.

In the Hall resistance (not shown)
two additional non-quantized plateaus are observed at fields slightly higher
than those of the commensurability peaks in $\rho_{xx}$. Around $B=0$ the
Hall effect is quenched.

\begin{figure}
\includegraphics{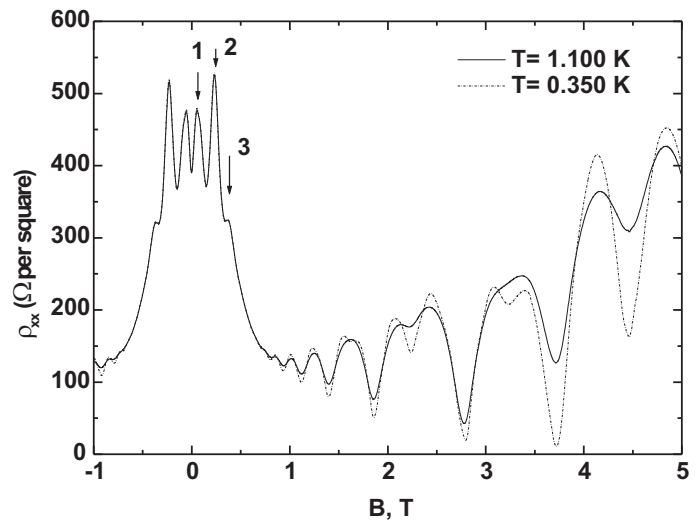} \caption{Magnetoresistance in sample D18-01-10
with a $700$~nm lattice period. One can see two main
commensurability peaks corresponding to ($\bf 1$) $r_c/a\approx
1.75$ and ($\bf 2$) $r_c/a\approx0.53$ around zero field and a
shoulder-like feature ($\bf 3$) at $r_c/a\approx0.33$.} \label{f.1}
\end{figure}

  The 700 and 850 nm samples are weakly sensitive to illumination which does
not change noticeably the electron density and results only in a slight
decrease of the zero field resistance. There is a weak temperature
dependence of the commensurability oscillations below $T=4.2$~K. On the
whole, apart from the shoulder, the behavior of the Si/SiGe samples with the
antidot lattice periods 700 nm and 850 nm is found to reproduce the typical
features well known from previous studies of GaAs/AlGaAs samples.

  The situation, however, is different in the samples with the superlattice
period 600 nm. The behavior of two such samples has been investigated. Fig.2a
shows the characteristic MR curves obtained for sample $D18-01-07$ after
cooling down and without prior illumination. Although the zero field
resistance ($\rho_{xx}(B=0)\approx 23$~k$\Omega$) is much higher than that of
the 700 and 850 nm samples, the electron density estimated from the slope of
the Hall resistance dependence is $N_s\approx 5.35\times10^{11}$cm$^{-2}$,
practically the same as in $D18-01-10$ sample, Fig.1.

  The peak $wl$ at zero magnetic field is due to the suppression of the weak
localization correction. The broad peak $\bf (1:2)$ at $B=0.21$~T
($r_c/a\approx0.64$) is supposed to be a superposition of the two main
commensurability peaks similar to peaks $\bf 1$ and $\bf 2$ in Fig.1. Normally,
no more commensurability features are expected at higher magnetic fields
before the ShdH oscillations set in.

  Nevertheless, in our 600 nm samples we observe at least three more peaks
$\bf 3,4,5$ at $B\approx 0.49$ ($r_c/a\approx0.28$), $0.72$
($r_c/a\approx0.2$) and $0.95$~T ($r_c/a\approx0.15$) respectively. The
variation of the peaks $\bf 3,4,5$ with temperature in Fig.2a is of the same
order as that of the commensurability peak $\bf (1:2)$ showing them to be of
a classical origin as well.

  Fig.2a shows a high resistance state of a 600 nm sample. However, it was
found that states of much lower resistance are possible in these samples
with the new peaks still present. Fig. 3a shows the magnetoresistance of our
second 600 nm sample $D18-01-08$ that upon cooling down usually had a
relatively low resistance state. One can see that despite more than a
tenfold difference in zero field resistance the MR dependence in Fig.3a
displays all the peculiar features observed in Fig.2a. Moreover, the
position of the new peaks in magnetic field is the same as for the $D18-01-07$
sample. The corresponding electron density estimated from the Hall
resistance is $N_s\approx 5\times10^{11}$cm$^{-2}$ which is slightly less
than that for sample $D18-01-07$ in Fig.2a. However, a precise determination
of the electron density is difficult due to the deformation of the Hall
resistance dependence in the 600 nm samples.

\begin{figure}
\includegraphics[width=6cm,height=8cm]{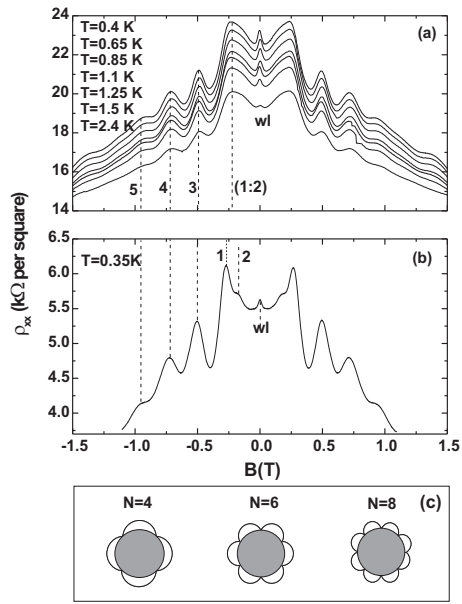}
\caption{Magnetoresistance of sample $D18-01-07$ with the lattice
period $600$~nm $\bf (a)$-before and $\bf (b)$-after illumination.
In $\bf (a)$ temperature increases from the upper to the lower
curve. The dashed lines mark the position of the weak localization
peak $\bf wl$, the commensurability peaks $\bf 1,2$ and the new
peaks $\bf 3,4,5$. $\bf (c)$-schematic drawing of periodic skipping
orbits with four, six and eight reflections from the antidot.}
\label{f.2}
\end{figure}

\begin{figure}
\includegraphics[width=6cm,height=8cm]{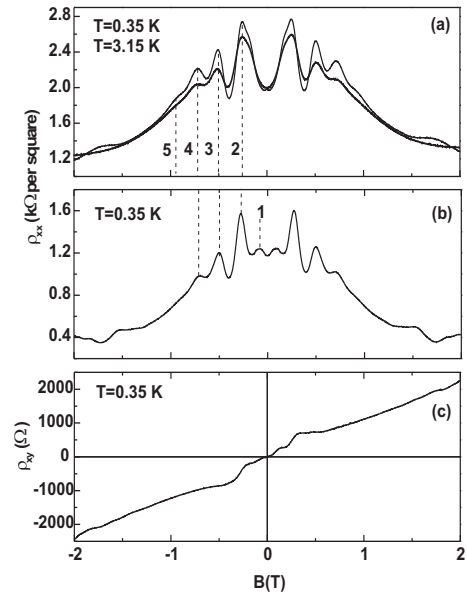}
\caption{Magnetoresistance of sample $D18-01-08$ with the lattice
period $600$~nm $\bf (a)$-before and $\bf (b)$-after illumination.
The numeration of the peaks is the same as in Fig.2. $\bf (c)$-Hall
resistance for the sample state after illumination $\bf (b)$. }
\label{f.3}
\end{figure}

  Fig.2b and 3b show the effect of illumination on the 600 nm samples. In
both samples illumination results in a decrease of zero field resistance. At
the same time within the experimental accuracy the electron density before
and after illumination remains practically the same. Therefore, the change
in resistance must be brought about by a mobility enhancement. This
conclusion is also supported by the fact that in both samples the broad peak
observed in the higher resistance state resolves itself after illumination
into two conventional commensurability peaks $\bf 1$ and $\bf 2$, Fig.2b,3b.
The position of peaks $\bf 3,4,5$ does not change with illumination.

  The Hall resistance dependence in 600 nm samples is practically the same
as in 700 and 850 nm samples with the quenching of the Hall effect and the
additional plateaux corresponding to the commensurability peaks (see
Fig.3c). No new features corresponding to the peaks $\bf 3,4,5$ in
$\rho_{xx}(B)$ have been found in the $\rho_{xy}(B)$ dependence.

  To our knowledge there have been at least two experimental studies
\cite{M.V.Budantsev,Eroms,Z.D.Kvon} of magnetotransport in antidot superlattices where
additional longitudinal resistance resonances were reported in magnetic
fields following the last comensurability peak $2r_c=a$ and before the onset
of Shubnikov-de Haas oscillations. In these experiments the lattice was
formed by large diameter antidots which seems to be crucial for the
observation of the additional peaks. The single resonance observed in
Ref.~\cite{M.V.Budantsev,Z.D.Kvon} was supposed to be due to a
localized electron trajectory locked in the free space pocket
between four adjacent antidots. To describe this resonance a simple commensurability condition
was suggested, \cite{M.V.Budantsev,Z.D.Kvon}: $2r_c=\sqrt{2}a-d_e$.
The shoulder-like feature observed in our $700$~nm samples
(peak $\bf 3$ in Fig.1) is likely to be of this kind.
Indeed, applying to it the mentioned commensurability condition
we obtain a reasonable value for the antidot effective diameter: $d_e\approx 540$~nm.
This model, however, predicts only a single additional resonance and it fails in
our $600$~nm samples where a whole series of oscillations is observed after
the main commensurability peak $2r_c=a$. In this case a more appropriate
approach is the one used in Ref.~\cite{Eroms} where two additional broad peaks
were reported.

The standard analysis performed in Ref.~\cite{Eroms}
included evaluation of the Kubo formula that yields
simulated magnetoresistance traces featuring the new peaks followed with the
examination of Poincare sections allowing the identification of the
electron trajectories responsible for the observed MR peaks.
It has been shown that after the cyclotron radius becomes smaller than the
distance between neighboring antidots a new class of electron trajectories
becomes important that under certain conditions can give rise to a new type
of magnetoresistance oscillations. In square lattices the new MR resonances
are thought to arise due to rosette-shaped orbits that
encircle the antidots and are localized similar to the pin-ball
orbits responsible for the commensurability
peak at $2r_c=a$. Also, it appears that contribution of such orbits to
magnetoresistance depends on whether they are periodic, i.e. reproduce
themselves on each revolution around an antidot, or non-periodic. It is
the periodic rosette-shaped orbits, like those shown in Fig.2c, that give rise
to individual maxima in magnetoresistance.

  In our experiment the new peaks $\bf 3,4,5$ lie at magnetic fields
  equal to $0.49$,
$0.72$ and $0.95$~T respectively. Assuming for simplicity the periodic
skipping orbit of order $N$ to be composed of $N$ semicircles, we obtain the
following simple geometric relation: $2r_c=d_{e}sin(\pi/N)$, where $d_{e}$
is the antidot effective diameter. Hence, $B_N \sim sin(\pi/N)^{-1}$ is the
magnetic field at which a magnetoresistance maximum corresponding to the
periodic skipping orbit of order "N" is to be expected. Now, it is easy to
show that the experimental maxima positions given above relate to each other
in approximately the same way as $B_N$ values corresponding to $N=4,6$ and
$8$. Indeed, multiplying $sin(\pi/N)^{-1}$ for $N=4,6$ and $8$ by the
same factor $0.346$, we obtain $0.49$, $0.7$ and $0.91$ in excellent
agreement with the experimental maxima positions. So, we attribute the peaks
$\bf 3,4,5$ in Fig.2,3 to the periodic skipping orbits of order $N=4,6$ and
$8$, shown in Fig.2c. This common factor $0.346$ also gives us the antidot
effective diameter $d_e$ in the $600$~nm samples. Indeed, taking
$N_s=5.35\times10^{11}$cm$^{-2}$ we
obtain $d_e\approx 490$~nm, which turns out to be more than two times larger
than the lithographical size of the antidots. This value is also close to the
effective diameter estimated from the position of the shoulder-like
feature in the $700$~nm period sample.

  It is worth mentioning that the oscillations $\bf 3,4,5$ observed in this
work are very well developed and comparable in size to the main
commensurability peak, far exceeding in this respect the features of this
type observed so far. Moreover, comparing our
results to \cite{Eroms}, where the experimental geometry was similar to
ours, we find that whereas we resolve three peaks, identified as
corresponding to $N=4,6$ and $8$, only two peaks $N=4$ and $N=8$ are
reported in \cite{Eroms} with peak $N=6$ missing not only experimentally
but in the calculation results as well. This and other facts raise a number
of questions that we shall list here. First, it is still not very clear
physically why periodic skipping orbits should be more efficient in trapping
electrons around an antidot and thus giving rise to maxima in
magnetoresistance than non-periodic skipping orbits. Also, it remains
unclear why only periodic skipping orbits of a certain order $N$ should give
rise to maxima in magnetoresistance, while others, say, $N=5$ or $N=7$,
should have no such effect. Finally, as seen when comparing our results with
Ref.~ \cite{Eroms}, the order $N$ of peaks that one will
observe seems to depend on particular experimental conditions. A possible
explanation to the latter questions may be that the sample specific
potential relief surrounding an antidot may be favorable to the formation
of skipping orbits of certain orders N only.

  Another important issue concerns our understanding of the antidot lattice
parameters that favor the observation of the new oscillations. So far, it
has been established that these oscillations can only be observed in
lattices with a sufficiently large ratio $d_e/a$, although it is not quite
clear why this should be important in the interpretation given above. In our
case, these oscillations are present when $d_e/a\approx 0.8$ and are already
absent in samples with $d_e/a \approx 0.7$. At the same time, in
Ref.~\cite{Eroms}, the oscillations are still observed in samples with
$d_e/a\approx 0.7$. On the other hand, the steepness of the antidot potential
does not seem to be an important factor. The new oscillations are present
both in InAs/GaSb samples where the antidot effective diameter is almost
equal to its lithographic size, \cite{Eroms}, and in our Si/SiGe samples
where, due to the depletion effects, the effective diameter is more than two
times larger than the lithographic diameter.

  To conclude, we have investigated the transport properties of square
antidot superlattices fabricated on Si/SiGe heterostructures. In lattices
with a comparatively small ratio $d_e/a$ ($d_e$ is the antidot effective
diameter, $a$ is the lattice constant) we observe the usual commensurability
maxima. In lattices with $d_e/a\approx 0.8$, apart from the usual
commensurability peaks, we observe three well developed
magnetoresistance resonances at higher magnetic fields. We attribute the new
resonances to a successive formation of rosette-shaped orbits encircling
an antidot and reflected from its boundary 4, 6 and 8 times, respectively.

\acknowledgments

This work has been supported by RFBR project no. 06-02-16129 and by ANR
PNANO MICONANO .

\end{document}